\documentclass{ametsocV6.1}
\usepackage{graphicx} 
\usepackage{float}
\usepackage{comment}

\title{Quantifying Very Extreme Precipitation and Temperature Using Huge Ensembles Generated by Machine Learning-based Climate Model Emulators}
\authors{Christopher J. Paciorek\correspondingauthor{Christopher J. Paciorek, paciorek@stat.berkeley.edu}\aff{a} and Daniel Cooley\thanks{}\aff{b}}

\affiliation{\aff{a}Department of Statistics, University of California, Berkeley, California, United States\\ \aff{b}Department of Statistics, Colorado State University, Fort Collins, Colorado, United States}

\abstract{Weather extremes produce major impacts on society and ecosystems and are likely to change in likelihood and magnitude with climate change. However, very low probability events are hard to characterize statistically using observations or even climate model output because of short records/runs. For precipitation, consideration of such events arises in quantifying Probable Maximum Precipitation (PMP), namely estimating extreme precipitation magnitudes for designing and assessing critical infrastructure. A recent National Academies report on modernizing PMP estimation proposed using very large climate model-based ensembles to estimate extreme quantiles, possibly through machine learning-based ensemble boosting. Here we assess statistical aspects of such an approach for the contiguous United States using a huge ensemble (10560 years) produced by a state-of-the-art emulator (ACE2) trained on ERA5 reanalysis. The results indicate that one can practically estimate very extreme precipitation and temperature quantiles, provided one uses appropriate statistical extreme value techniques. More specifically, the results provide evidence for (1) the use of threshold-exceedance methods with a sufficiently high threshold (necessary for precipitation) for reliable estimation, (2) the robustness of results to variation in extremes by season and storm type, and (3) the sufficiency of the ensemble for well-constrained statistical uncertainty. Our results also show that the emulator produces extremes outside the range of the ERA5 training data. While encouraging for emulators' potential use for quantifying the climatology of extremes, more investigation is needed to assess whether emulators are fit for this purpose. Our focus is on how to use huge ensembles to estimate very extreme statistics; we expect the results to be relevant for future improved emulators.}

\begin{document}

\maketitle 
\statement   
Weather extremes produce major impacts on society and ecosystems and are likely to change in likelihood and magnitude with climate change. Machine learning now provides methods to simulate weather for many hypothetical years at much lower computational expense than traditional models. This could provide the opportunity to estimate the magnitude of very low probability events using statistical methods applied to the simulation output, for use in planning, risk assessment, and adaptation. Our work finds that when used appropriately, statistical methods designed to analyze extremes can reliably estimate the magnitude of extremely high precipitation and temperature events in the United States, under the critical assumption that the machine learning method is a good mimic of real world weather.  

\capsule 
Statistical analysis of 10560 years of simulated weather from a machine learning-based climate model emulator supports an approach to estimating the magnitude of very extreme precipitation and temperature events.

\section{Introduction}

Estimating the magnitude of very extreme weather events is essential for infrastructure design, 
risk assessment, and emergency preparedness.  
Impacts from extreme events can occur through flooding, heatwaves, drought, high winds, and other mechanisms.
In this work, we are specifically interested in very low probability events, ones whose annual exceedance probability (AEP) could be orders of magnitude smaller than the reciprocal of the length of the observation record.
For example, we may aim to estimate the magnitude of a 1-in-10000-year event (e.g., the 1-in-10000-year AEP precipitation depth), or an event with even lower probability. (AEP values are also referred to as ``return values" or ``return levels".)
The short sample sizes available from observations or computationally-expensive climate models are inadequate to provide direct location-specific empirical estimates of such events. 


An important use case for estimating such extreme events is PMP, a long-standing concept for quantifying very extreme precipitation for use in designing and assessing the safety of critical infrastructure such as dams and nuclear power plants. 
PMP has historically been defined as the maximum possible precipitation over a given area and time duration \citep{nasem2024pmp}.
In the U.S., PMP has been estimated using an approach that combines 
a catalog of the most extreme observed events with storm transposition (a technique for trading space for time by judging where else an observed storm could plausibly occur), moisture maximization, and orographic adjustment. 
While motivated by PMP, we also consider temperature.

\cite{nasem2024pmp} proposed a new approach to PMP. 
In place of trying to estimate the maximum possible precipitation, NASEM proposed a new definition of PMP, ``the depth of precipitation for a particular duration, location, and areal extent, such as a drainage basin, with an extremely low annual probability of being exceeded, for a specific climate period".   
In conjunction with this new definition, and in recognition of methodological and conceptual difficulties with PMP in practice, NASEM recommended the use of climate models
to generate very large ensembles and estimation of PMP as quantiles of the ensemble-based precipitation distribution (for any desired spatial or temporal aggregation) using statistical extreme value analysis. 

A statistical extreme value analysis (EVA) typically aims to characterize the magnitude of extreme events, often beyond the range of the data.
At EVA's foundation is probability theory that characterizes the limiting distributions of random variables as they increase toward their upper limit (infinity if unbounded) \citep{coles2001introduction}.
In practice, an EVA begins with the researcher taking a small subset of the most extreme data values from the full data (typically much less than 5\%). This subset is considered to provide the best information about the distribution's tail, whose characterization is the critical component for estimating the magnitude of very low probability events. 
Two general approaches are widely used: analysis of block (e.g., annual) maxima or of threshold exceedances. 
Even though EVA was specifically designed to estimate far into a distribution's tail, 
it can still be difficult to use EVA to obtain meaningful estimates of very low probability events.
Simply put, it is difficult to take $\sim$100 years of data and say something relevant about a 1-in-10000 year event. 
When extrapolating far into the tail, EVA can provide a  statistical estimate, but large uncertainty can limit its usefulness.


Traditional climate and meteorological models are computationally expensive. Increased computing power has drastically increased our ability to run models at higher resolution and with better physical fidelity, and to create large ensembles to try to characterize uncertainty and variability \citep{FlatoEtAl2013,nasem2024pmp}. Coupled with this has been a long history of developing stochastic weather generators \citep{ailliot2015stochastic}, climate model emulators \citep{tebaldi2025emulators}, regional climate models \citep{giorgi2019thirty} and other techniques to produce simulated weather comparable to (or at higher resolution than) that from models, but with less computation. Recent advances in machine learning (ML) have resulted in very fast emulators that perform better than traditional approaches \citep{bi2023accurate,lam2023learning,price2025probabilistic}. At the time of \citet{nasem2024pmp}, ML-based emulators were used primarily at weather time scales for forecasting, but there are now promising climate model emulators \citep{chapman2025camulator,cresswell2024deep,watt2025ace2}. This opens the possibility of generating huge ensembles and statistically analyzing the output to characterize extremes. 






While \cite{nasem2024pmp} proposed a new definition and a climate model-based approach for estimating PMP, it expected that development of the new approach would occur over a period of years. The report anticipated that ensemble boosting might play a key role in increasing ensemble sizes. Indeed, less than two years after report publication, for this work we produced a huge ensemble with little difficulty. In particular, we use ACE2, for which extensive validation work has been done \citep{watt2025ace2} and for which code and input data are readily available. With this ensemble, we can investigate the statistical techniques one could use to analyze extremes using huge ensembles. Tying that investigation loosely to the NASEM recommendations, we note that the report anticipated that to avoid bias from inclusion of non-extreme precipitation values when fitting statistical distributions, one would want to use either very high thresholds  for threshold exceedance analysis or very long blocks for block maxima-based analysis. 
A threshold-based approach can naturally exclude events from storm types or seasons that do not produce very extreme precipitation. Finally, the report presented sample size calculations showing that the sample size needed to reduce uncertainty sufficiently depended critically on the tail of the distribution. For tails that are not ``too" heavy, one might need on the order of 1000 exceedances of the chosen threshold.

In this work we investigate the following key questions regarding extreme precipitation and temperature:

\begin{itemize}
\item Can the emulator produce very extreme values (an initial, limited assessment of whether the emulator is fit for purpose)?
\item  Given the importance of tail behavior for quantifying extremes, what is the tail behavior of the emulator? Is there evidence for heavy-tailed precipitation distributions?
\item Do extreme value methods provide a reasonable approach to estimating very low probability AEP values?
\item Can one limit bias in AEP value estimation by using only the most extreme values, without unduly increasing statistical uncertainty?
\item Can an appropriate threshold exceedance-based analysis avoid the difficulties of storm typing or seasonal analysis?
\item What sample sizes are required for reasonable statistical certainty?
\end{itemize}

  Using an ML-based emulator trained on reanalysis provides a testbed to assess the recommendations of \citet{nasem2024pmp} empirically and give guidance for statistical analysis of huge ensembles beyond the PMP use case.
For real applications in the future, there are key advantages to using suitably-trained ML-based emulators:  large sample sizes, the potential to produce realistic space-time fields, and the potential to produce out-of-sample events by training on the observed evolution of the earth system.

\section{Methods}

\subsection{ACE2-ERA5 ensemble}

We generated a huge ensemble using the ACE2 ML-based climate model emulator, specifically ACE2-ERA5, trained on ERA5 reanalysis data \citep{watt2025ace2}. ACE2 is an autoregressive emulator that operates at 1-degree spatial resolution and 6-hour temporal resolution, with eight vertical layers in the atmosphere. \cite{watt2025ace2} provides details, including extensive validation work, and shows that the emulator produces stable simulation for arbitrarily long time intervals.

We created an ensemble with a total of 10560 years of emulated weather, based on 12 initial conditions, run for 22 years of forcing data, and 40 repetitions (details in Appendix A). We aggregate to the daily scale by averaging the surface precipitation rate (before converting to precipitation per day) and taking the maximum 2m air temperature. For analysis we restrict to the grid cells contained in the contiguous U.S. 

\subsection{Statistical extreme value analysis}


We use the climextRemes package \citep{climextRemes} to carry out EVA separately for each grid cell, fitting both threshold exceedance and block maxima models using maximum likelihood. 
For block maxima analysis, we fit the generalized extreme value (GEV) distribution to annual maxima.
For the threshold exceedance (peaks-over-threshold [POT]) analysis, we fit a point process representation of the generalized Pareto distribution (GPD) to observations exceeding a threshold \citep[Section 7]{smith1989extreme,coles2001introduction}.
Parametrizations for the GEV and GPD are given in Appendix A and correspond to \cite{coles2001introduction}.
GPDs are fit to a series of increasing thresholds, corresponding to the upper 0.001 to 0.00001 quantiles of daily observations, spaced on a regular grid of 10 quantiles on the log scale. 
These correspond to selecting the largest 3857, 2313, 1387, 831, 499, 299, 180, 108, 65, and 39 observations. 
AEP values are estimated using the parameter point estimates, with AEP value uncertainty calculated using the statistical delta method  on the MLE-based information matrix. 
For probabilities that are low but not too low, the size of our ensemble also allows us to assess EVA results by comparing AEP value estimates to empirical quantiles \citep[e.g.,][]{huang2016estimating}.

Our statistical fitting assumes independence across all years of the ensemble, which seems reasonable given the limited predictability of the atmosphere and that the ensemble's years encompass different starting values and different years of forcing data. 
Also we assume stationarity, in part given the use of only 22 years of forcing data but also for convenience \citep[see also][]{ben2020evaluation}. 
Our main goal is to explore big-picture statistical questions, rather than to account for finer-grained sources of variability. 
Our analysis treats the days within a year as independent, which could cause our estimation of uncertainty to be too low. 
There are widely-used techniques accounting for short-term dependence 
(e.g., declustering or bootstrapping) that can be used in practice. 
For simplicity, these are not considered here.
The analysis does not account for dependence between the grid cells. 
Some of this dependence is caused by true dependence in precipitation and temperature between grid cells (at both weather and climate time scales), while some is likely from the known spatial smoothing present in the ACE2 emulator \citep{watt2025ace2}.

Of particular interest will be the shape parameter $\xi$, which appears in both the GEV and GPD and which determines the upper tail's behavior.
If $\xi > 0$ the upper tail is unbounded and heavy, if $\xi = 0$ the upper tail is unbounded but light, and if $\xi < 0$ the tail is bounded.
Precipitation is often found to have a heavy tail \citep{nasem2024pmp}, while temperature is often found to be bounded \citep[e.g.,][]{philip2022rapid,bercos2022anthropogenic}.
We also investigate a bias-variance trade-off inherent to extremes.
Bias arises because the GEV or GPD become the correct distribution only as the block size increases to infinity or as the threshold approaches the distribution's upper limit.
As any data set is finite, increasing the block size or threshold decreases the number of retained observations, thereby increasing the uncertainty in estimating parameters or quantiles.
The practitioner must decide how to balance bias and variance, and estimates can be sensitive to block length or threshold \citep[e.g.,][]{ben2020evaluation}. 

\subsection{Seasonality and storm types}

Seasonal effects imply weather data are not identically distributed.
Seasonality can be explicitly accounted for, either by performing separate analyses for each season or by building seasonality into the model's parameters.
On the other hand, because quantities of interest (e.g., AEP depths) are often on an annual time scale, seasonality is sometimes ignored.
For example, by setting the block length to correspond to one year, the analyst essentially looks at extreme behavior across all seasons.
However, if extreme observations occur only during a portion of the year, the effective block size $d$ is less than 365, and estimates could be biased due to inadequate block size.

A similar issue arises with storm types \citep{nasem2024pmp}.
Impactful events could arise from different weather phenomena; for example, extreme precipitation could arise from tropical cyclones or strong convective storms. 
Thus, one might consider the overall precipitation distribution to be best represented statistically as a mixture of distributions over the different storm types, or, as an approximation, as a mixture over different seasons, to the extent that the storm types in a season are a subset of the types seen through the year.
To account for the mixture, one could stratify the extreme data by type and perform separate analyses. 
Doing so fits data that are more homogeneous, better characterizing the distribution's tail, which is the key to extrapolation. 
The challenge with this approach is that this requires data labeled by storm type, and AEP estimation also requires estimation of the occurrence frequency of each storm type.
Alternatively, one could ignore storm types, thereby essentially aggregating extreme behavior across storm types.
The challenge with this approach is that aggregating over storm types carries the risk of bias when estimating the magnitude of very low probability events, especially if these most extreme events predominantly arise from a single storm type but the threshold exceedances include (presumably generally less extreme) values from other types.
To estimate PMP, \citet{nasem2024pmp} recommended a threshold exceedance approach applied to data unlabeled by storm type and advocated setting the threshold high enough to exclude most, if not all, of the data arising from storm types other than most extreme.
 



We explore the effect of fitting season-specific (DJF, MAM, JJA, SON) models under two approaches: (1) using the same thresholds as discussed above for each season (but excluding the two highest thresholds as these produce too-small sample sizes at the seasonal level) 
and (2) using the same number of exceedances in each season as in the full-year analysis. The former uses the same threshold and therefore smaller sample sizes (in some cases zero observations such as for temperature in winter in many locations), while the latter uses lower (or equal when all the annual exceedances occur in a single season) thresholds and larger sample sizes.

\section{Results}

\subsection{Precipitation}

We start by analyzing precipitation before turning to temperature.

\begin{figure}
    \centering
    \includegraphics[width=\linewidth]{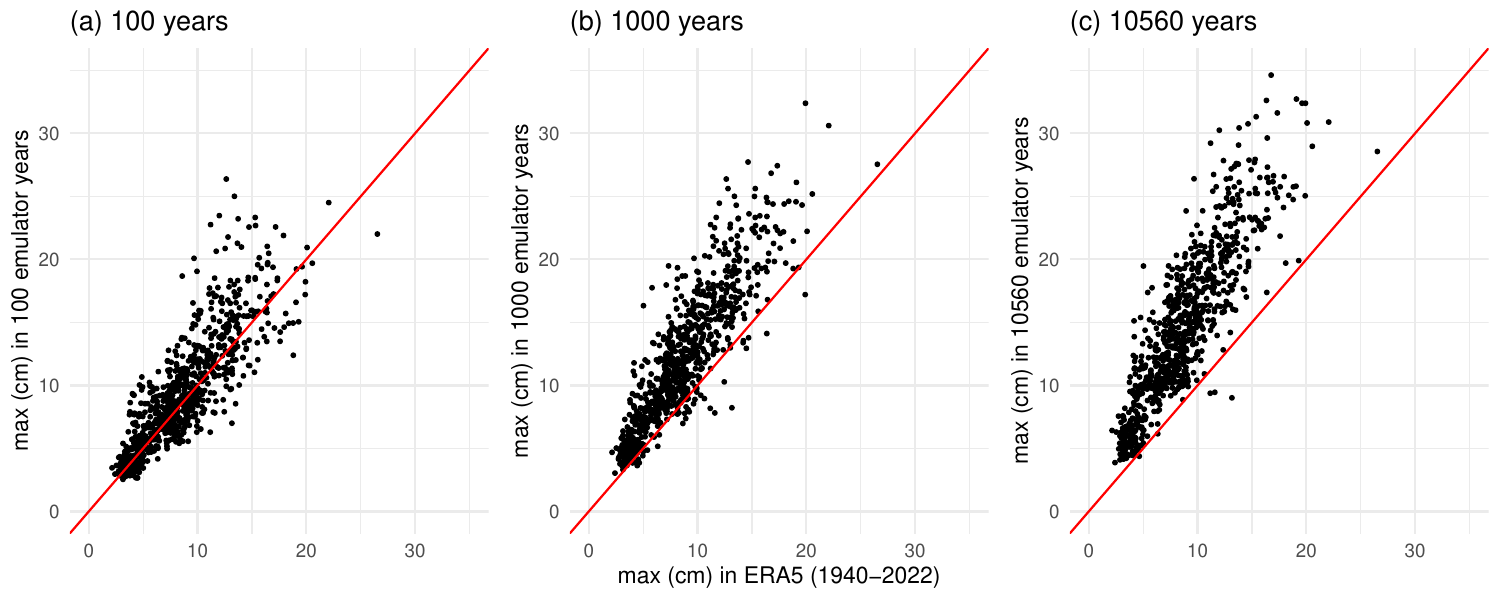}
    \caption{Maximum daily precipitation for each grid cell over the specified time period from emulator (for different numbers of simulated years) compared to maximum in ERA5 (1940-2022). One-to-one lines are in red.}
    \label{fig:prec-maxes}
\end{figure}

Fig. \ref{fig:prec-maxes} shows the maximum daily precipitation in each grid cell for different ensemble sizes compared to the maximum from ERA5 (83 years; of which most were used for training). The smallest ensemble size has maxima that generally correspond to the ERA5 data, with some variability as expected from running a stochastic emulator. As the ensemble size increases, the location-specific maxima generally increase and exceed the ERA5 maxima, indicating that the emulator produces out-of-sample extremes relative to the training data. Full exploration of the quality of the emulator relative to the true climate system is beyond the scope of this work and is difficult for extremes. However, the ability of the emulator to extrapolate into the tail of the distribution suggests that the precipitation extremes in an ensemble could be useful for understanding extremes in the climate system.

\begin{figure}
    \centering
    \includegraphics[width=\linewidth]{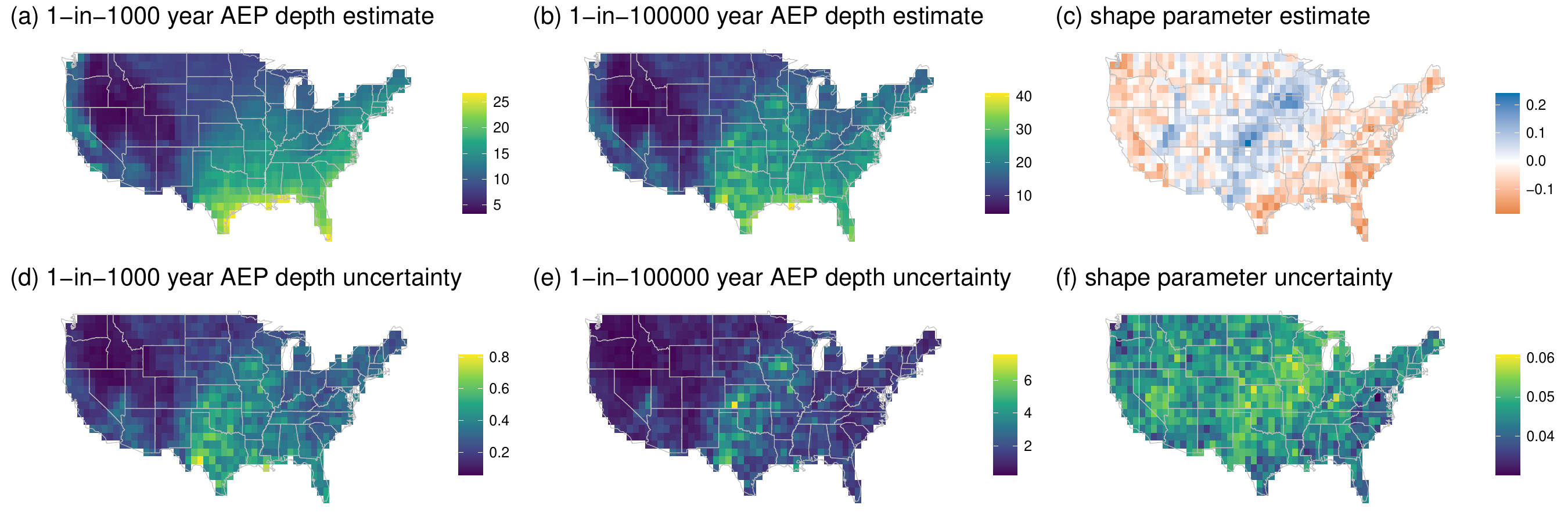}
    \caption{Emulator-based extreme precipitation climatology: (a) 1-in-1000 year AEP depth (cm) estimates, (b) 1-in-100000 year AEP depth (cm) estimates, (c) shape parameter estimates, with corresponding standard error estimates (d), (e), and (f), respectively.}
    \label{fig:prec-climate}
\end{figure}

Fig. \ref{fig:prec-climate} provides information about the extremes climatology (estimated AEP depths and shape parameters) in the ensemble, based on the threshold exceedances approach using $n$=499 exceedances.
The AEP depth estimates show spatial patterns in accordance with U.S. precipitation climatology: the most extreme precipitation is seen on the Gulf Coast and the least extreme precipitation is in the dry interior of the western U.S. The spatial pattern in the shape parameter estimates is less intuitive, but one interesting feature is that many of the estimates (56\%) are negative (corresponding to bounded distributions), in contrast with previous analyses that generally estimate positive shape parameters (summarized in \citet{nasem2024pmp}). 
We explore this result further below. 

\begin{figure}
    \centering
    \includegraphics[width=\linewidth]{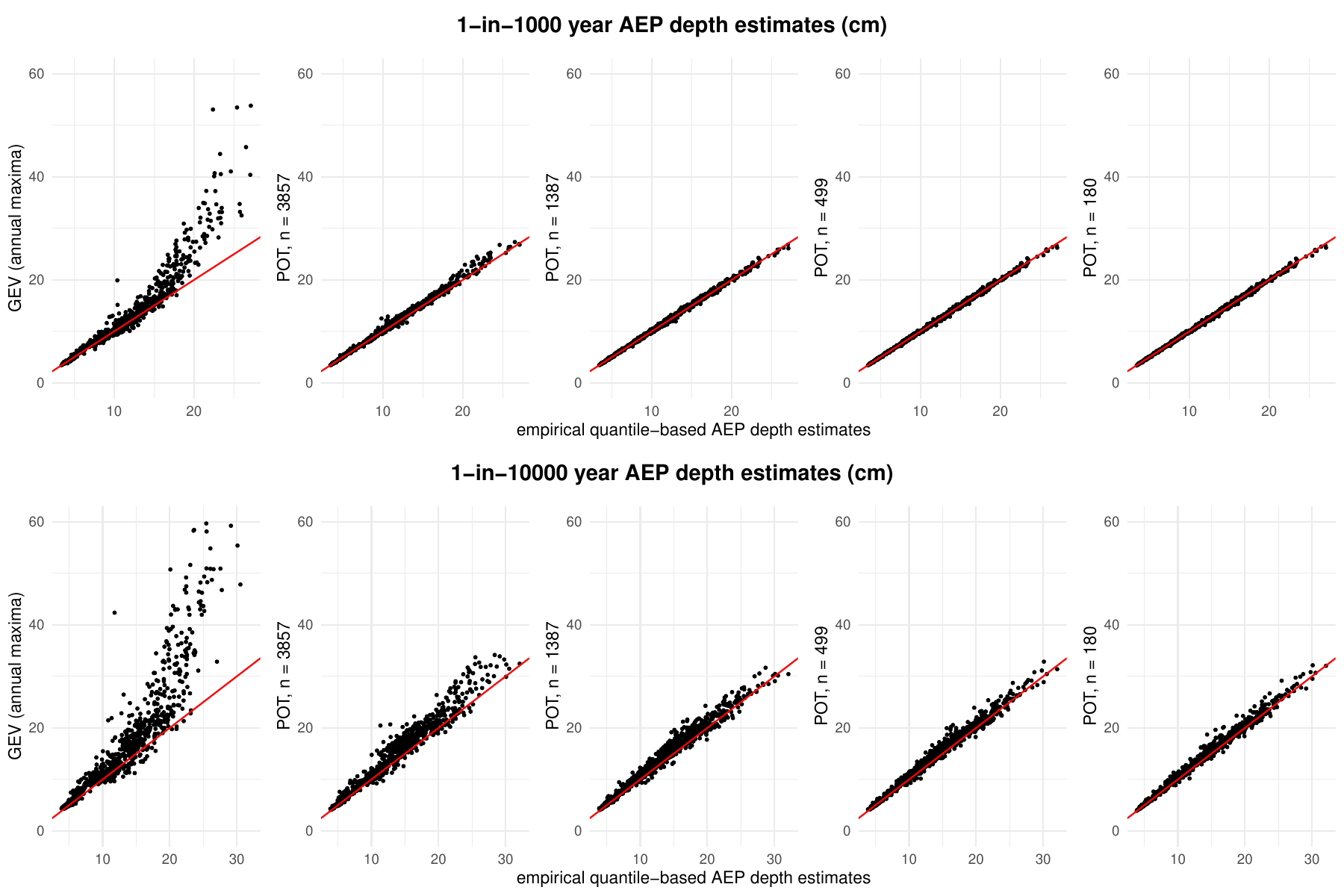}
    \caption{AEP depth estimates for each grid cell from annual maximum-based GEV fit (first column) and POT fits (remaining columns, with increasing thresholds from left to right) compared to AEP depth estimates from empirical quantiles, with 1-in-1000 year estimates in top row and 1-in-10000 year in bottom row. One-to-one lines are in red.}
    \label{fig:prec-rv-bias}
\end{figure}

Next we assess the performance of the GEV and POT fitting approaches for estimation of AEP depths corresponding to shorter time periods, 1000 and 10000 years, that are within the range of our ensemble size. By investigating shorter periods, we can estimate AEP depths directly as empirical quantiles of the samples in the ensemble and use them to assess the performance of EVA-based AEP depth estimates. Fig. \ref{fig:prec-rv-bias} shows that when the POT threshold is sufficiently high (e.g., using sample sizes of 499 and fewer), the AEP depth estimates are very similar to the empirical quantiles. In contrast, for lower thresholds (larger sample sizes), there appears to be a modest upward bias. In stark contrast, GEV estimation based on annual maxima (the most common choice in practice) appears to be severely upwardly-biased, particularly for grid cells with more extreme precipitation.
The fact that POT-based AEP depths are well-aligned with empirical quantile-based estimates for shorter time periods provides support for using POT-based AEP depth estimates for longer time periods (such as 100000 and one million years). Block maxima with blocks of a year are clearly insufficient for good estimation, as is also the case for POT estimation using the same sample size (results not shown). 
Use of longer blocks (e.g., 5, 10, 20, ... years) for GEV analysis produces results (not shown) equivalent to those seen here for POT, but here we focus on POT methods and sensitivity to the threshold.

\begin{figure}
    \centering
    \includegraphics[width=\linewidth]{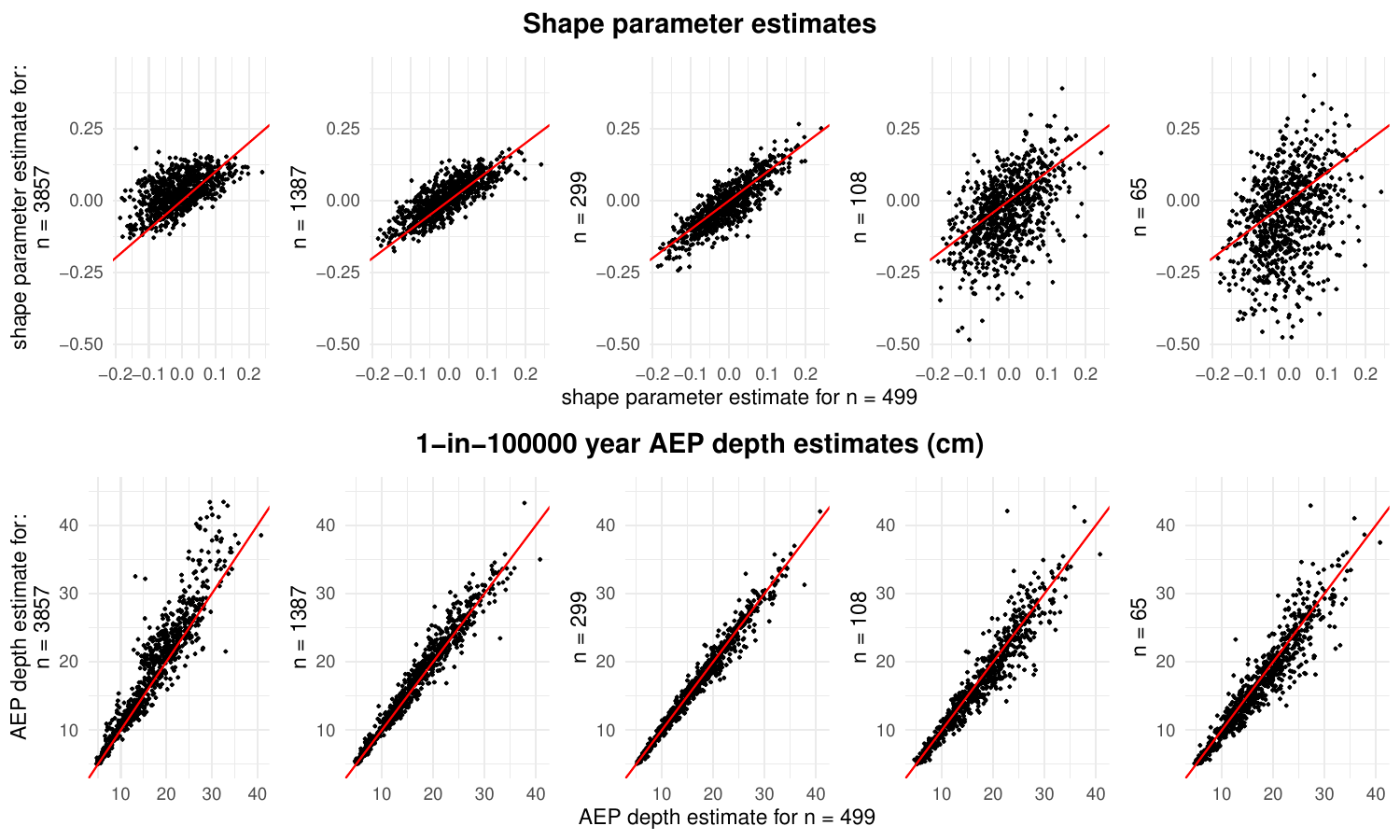}
    \caption{Change in estimates with threshold for POT-based estimates for each grid cell. (Top row) Shape parameter estimates with increasing threshold (decreasing sample size) on y-axis, compared to estimates with $n=499$ on x-axis. (Bottom row) 1-in-100000 year AEP depth estimates with increasing threshold, compared to estimates with $n=499$. One-to-one lines are in red.}
    \label{fig:prec-stability}
\end{figure}

We next explore the tail behavior of precipitation, in relation to the choice of threshold. Fig. \ref{fig:prec-stability} (top row) shows that shape parameter estimates systematically decrease (tending to move from above the 1:1 line to below it) as one increases the threshold (decreases the sample size of exceedances), with variability also increasing, particularly dramatically with small sample sizes. Fig. \ref{fig:prec-shape-hists} shows that the number of positive shape parameter estimates decreases substantially, with the distributions shifting to the left (as well as becoming more variable with smaller sample sizes). Correspondingly, AEP depth estimates decrease as the threshold increases (Fig. \ref{fig:prec-stability}, bottom row). While the AEP depth estimates do decrease systematically, for the larger threshold values, they largely stabilize. From an implementation perspective, this suggests that if one chooses a sufficiently high threshold (such as corresponding to $n=499$), one can reasonably estimate AEP depths with limited bias. From a conceptual perspective, the decrease in shape parameter estimates suggests that previous results showing positive estimates for precipitation could be driven at least in part by the (necessary) use of low thresholds, potentially causing bias by including exceedances not representative of the far tail of the distribution. Of course this emulator-produced output could fundamentally differ from actual precipitation,  and the decrease in shape parameter estimates could be a result of the emulator output having a lighter tail than reality.

\begin{figure}
    \centering
    \includegraphics[width=\linewidth]{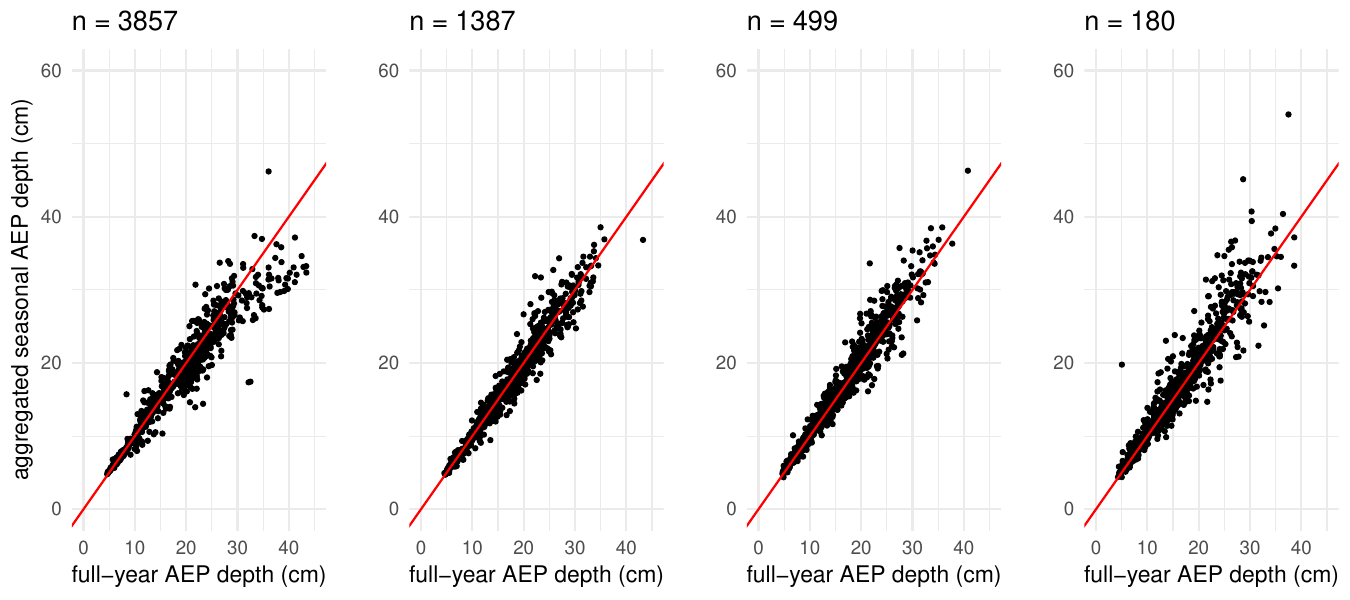}
    \caption{Comparison of 1-in-100000 year AEP depth estimates based on analysis of full-year data to aggregation over seasonally-stratified estimates for each grid cell. Results are shown for the second approach discussed in the methods; results for the first approach are similar (Fig. \ref{fig:prec-rv-seasonal1}). One-to-one lines are in red.}.
    \label{fig:prec-rv-seasonal2}
\end{figure}

The analyses above ignore seasonality and storm types. Fig. \ref{fig:prec-rv-seasonal2} shows that if one fits season-specific POT models and  then determines the overall AEP depth as the value at which the sum of exceedance probabilities across the four seasons equals the desired probability, the results are very similar to ignoring seasonality, provided the threshold is sufficiently high. This is consistent with extreme value theory, which requires a sufficiently high threshold to remove the influence of values from the bulk of the distribution. In the climate context, this removes smaller extremes corresponding to storm types (or seasons) that produce less extreme precipitation and naturally includes only the storm type(s) producing very extreme precipitation. From a practical perspective, being able to  avoid storm typing or the need to analyze seasons individually can be a helpful simplification, particularly avoiding the subjectiveness of and uncertainty from categorizing events into types.

\begin{figure}
    \centering
    \includegraphics[width=\linewidth]{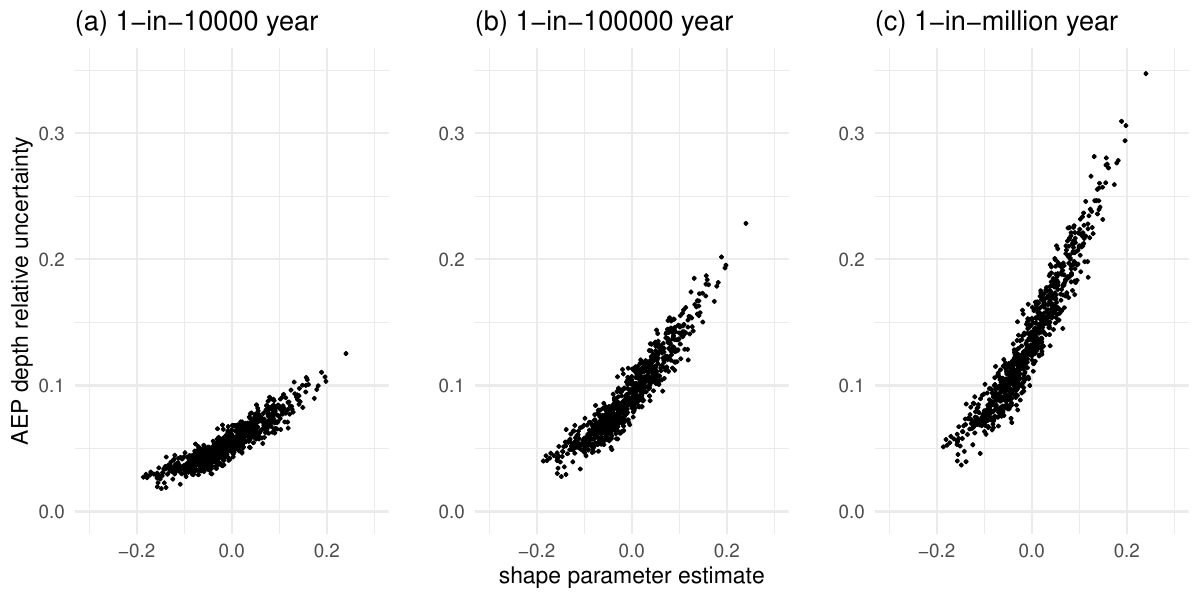}
    \caption{Relative uncertainty (standard error divided by estimate) for each grid cell for (a) 1-in-10000 year, (b) 1-in-100000 year, and (c) 1-in-million year AEP depth estimates as a function of the shape parameter estimates, using POT analysis with sample size of 499.}
    \label{fig:prec-rv-uncertainty}
\end{figure}

Finally, we turn from bias in estimation to focus on  statistical uncertainty, characterizing whether an ensemble of size 10560 is able to sufficiently reduce uncertainty in AEP depth estimates for practical use. As discussed in \citet{nasem2024pmp}, uncertainty is driven primarily by the sample size, the tail behavior, and the exceedance probability of interest. Fig. \ref{fig:prec-rv-uncertainty} shows the relative uncertainty (the standard error divided by the AEP depth estimate) for three time periods when using $n=499$ exceedances. Except when considering the longest period in combination with the most positive shape parameter estimates, relative uncertainty is generally less than 15\%, corresponding to confidence intervals whose length is plus or minus 30\% of the value of the estimate (e.g., an estimate of 20 cm with a confidence interval of (14, 26)). While the standard errors are approximate (see Methods), our purpose is to approximately characterize uncertainty without delving into the details of alternative methods (e.g., the bootstrap) for estimating uncertainty.

For many applications/locations, a 10560-member ensemble may be sufficient. However, if not, it is feasible now, and will become more so, to produce even larger ensembles, while cautioning that the benefit of this diminishes because relative statistical uncertainty decreases based on the inverse of the square root of the sample size.

\subsection{Temperature}

For temperature, we carry out a similar, but more briefly discussed, set of analyses.

\begin{figure}
    \centering
    \includegraphics[width=\linewidth]{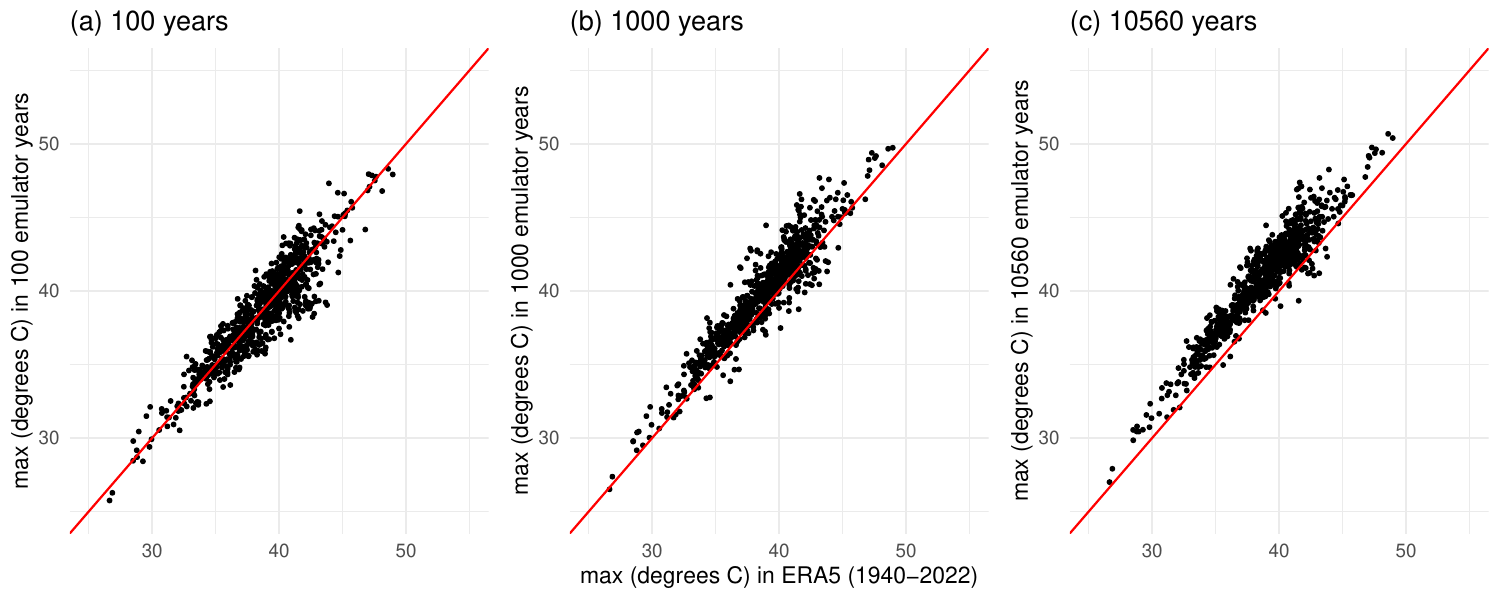}
    \caption{Maximum daily temperature for each grid cell over the specified time period from emulator for different numbers of simulated years compared to maximum in ERA5 (1940-2022). One-to-one lines are in red.}
    \label{fig:temp-maxes}
\end{figure}

Fig. \ref{fig:temp-maxes} shows the maximum daily temperature in each grid cell for different ensemble sizes compared to the maximum from ERA5. Results are similar qualitatively to precipitation, with increasing emulator size producing larger extremes, suggesting that the emulator can produce out-of-training-sample values. 

\begin{figure}
    \centering
    \includegraphics[width=\linewidth]{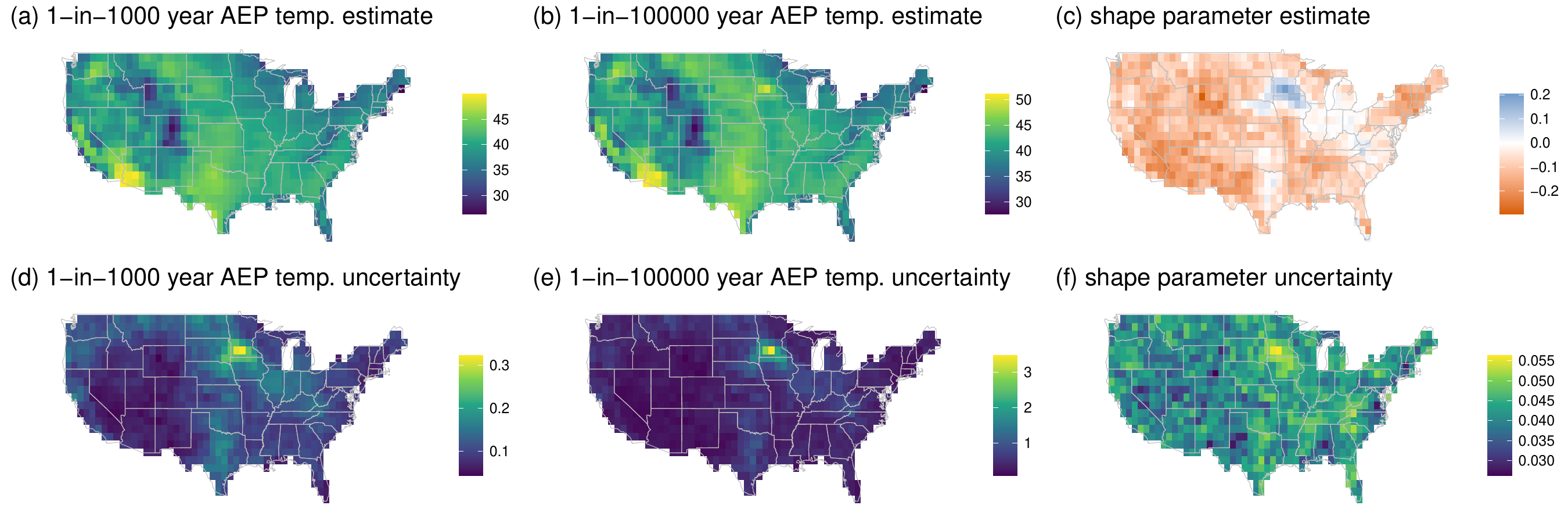}
    \caption{Emulator-based extreme temperature climatology: (a) 1-in-1000 year AEP temperature (deg.\ C) estimates, (b) 1-in-100000 year AEP temperature (deg.\ C) estimates, (c) shape parameter estimates, with corresponding standard error estimates (d), (e), and (f), respectively. The local maximum seen in the north-central U.S. (southern Minnesota), an area with little topographic variation, appears to be from sampling variability and not optimization problems and is associated with coinciding anomalously large shape parameter estimates.}
    \label{fig:temp-climate}
\end{figure}

Fig. \ref{fig:temp-climate} provides information about the extremes climatology (estimated AEP temperatures and shape parameters) in the ensemble, based on the threshold exceedances approach using $n=499$ exceedances. The AEP temperatures show spatial patterns expected based on U.S. temperature climatology with the most extreme temperatures seen in the more arid inland areas of the western U.S. As expected, the shape parameter estimates are almost all negative. 

\begin{figure}
    \centering
    \includegraphics[width=\linewidth]{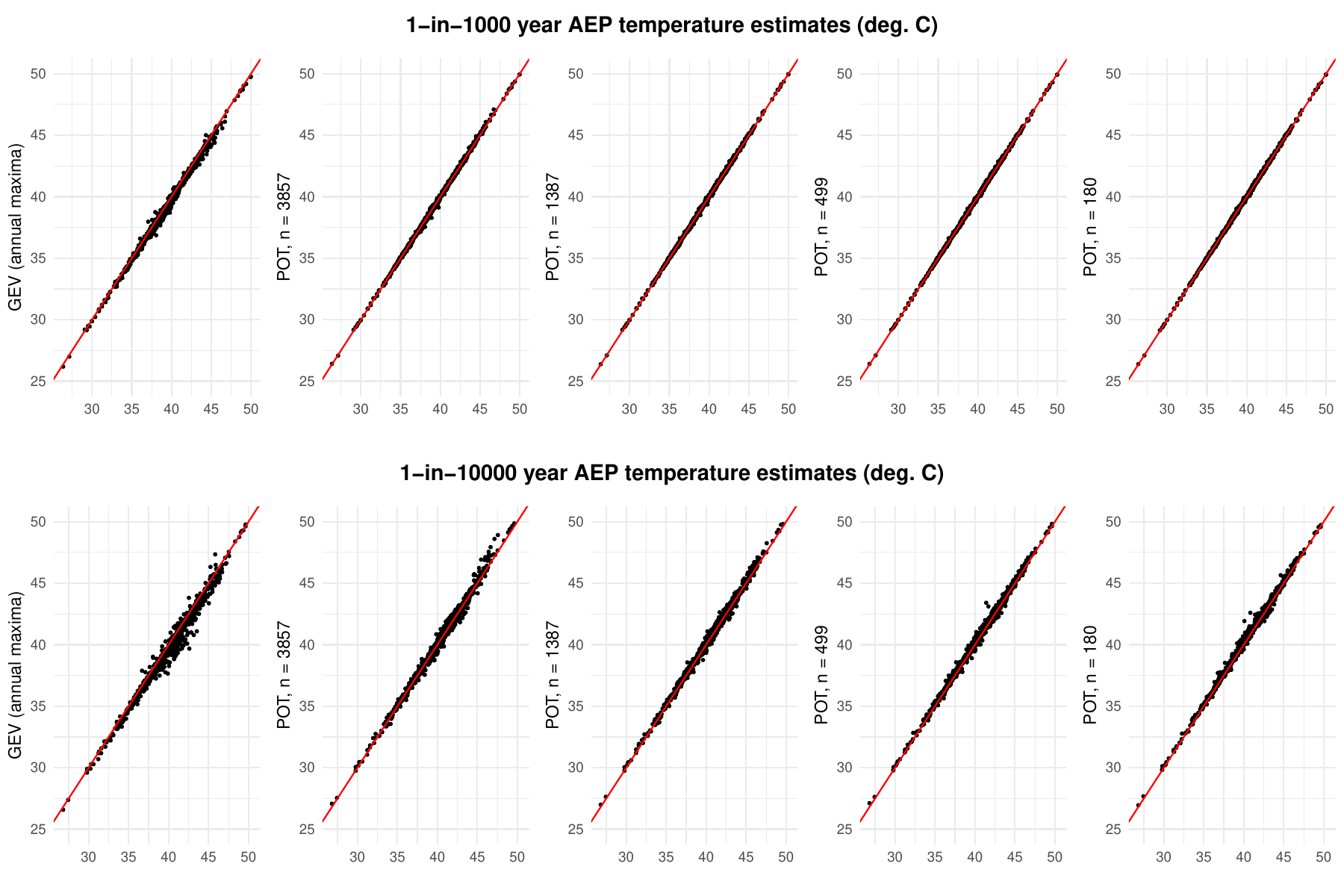}
    \caption{AEP temperature estimates from annual maximum-based GEV fit (first column) and POT fits (remaining columns, with increasing thresholds from left to right) compared to AEP temperature estimates from empirical quantiles, with 1-in-1000 year estimates in top row and 1-in-10000 year in bottom row. One-to-one lines are in red.}
    \label{fig:temp-rv-bias}
\end{figure}

Next we assess the performance of the GEV and POT extreme value fitting approaches for estimation for AEP temperatures corresponding to shorter time periods (1000 and 10000 years) to assess the performance of EVA. Fig. \ref{fig:temp-rv-bias} shows that both GEV and POT (regardless of the threshold) produce AEP temperature estimates that are very similar to the empirical quantiles. This contrasts with the bias seen for precipitation; with the more bounded temperature distributions, it is not necessary to use longer blocks or larger thresholds to minimize bias. These results provide support for using the EVA-based AEP temperature estimates for longer periods (such as 100000 and one million years). We did notice that the numerical optimization for the GEV approach was more fragile than for POT, requiring more careful initialization to find the maximum of the log-likelihood at many locations.

Fig. \ref{fig:temp-stability} shows that AEP temperature estimation is robust to the choice of threshold in the POT analysis, even for low thresholds (unlike for precipitation). This is good news in that use of lower thresholds, with more exceedances, reduces variance in estimation. This is consistent with the shape parameter estimates not showing a relationship with the threshold (apart from increasing variability with smaller sampler sizes) (Fig. \ref{fig:temp-shape-hists}).

\begin{figure}
    \centering
    \includegraphics[width=\linewidth]{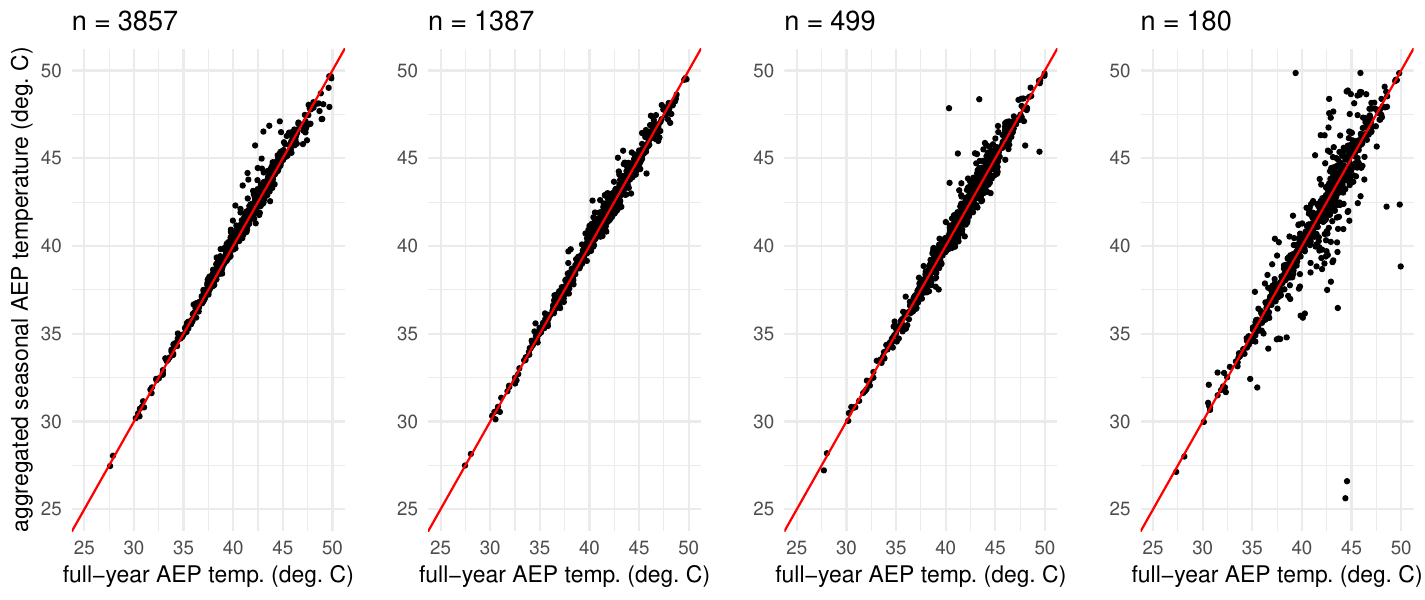}
    \caption{Comparison of 1-in-100000 year AEP temperature estimates based on analysis of full-year data to aggregation over seasonally-stratified estimates. Results are shown for the second approach discussed in the methods. One-to-one lines are in red.}
    \label{fig:temp-rv-seasonal2}
\end{figure}

Fig. \ref{fig:temp-rv-seasonal2} shows that, as for precipitation, one can achieve very similar results using EVA analysis applied directly to the full data compared to seasonally-stratified analysis, a helpful simplification. Those results use the second approach discussed in the methods. With the first approach, the full-year and maximum seasonal results are almost identical (Fig. \ref{fig:temp-rv-seasonal1}), because almost all of the seasonal maxima results are from the summer season fit to nearly the same set of exceedances as the full-year analysis.

\begin{figure}
    \centering
    \includegraphics[width=\linewidth]{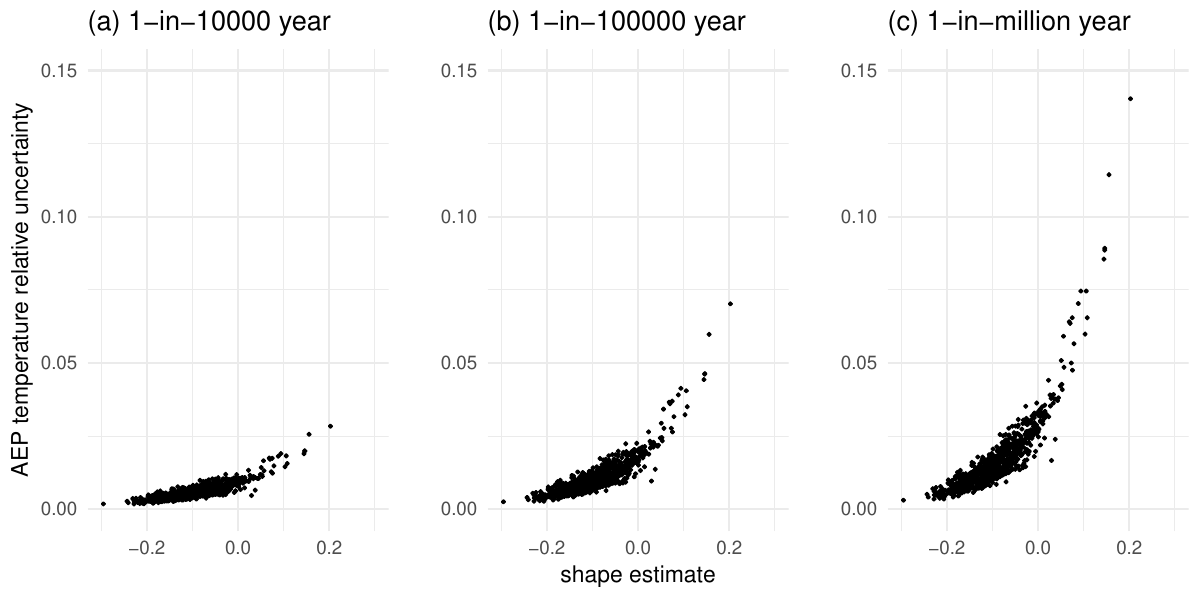}
    \caption{Relative uncertainty (standard error divided by estimate) for (a) 1-in-10000 year, (b) 1-in-100000 year, and (c) 1-in-million year AEP temperature estimates as a function of the shape parameter estimates, using POT analysis with sample size of 499.}
    \label{fig:temp-rv-uncertainty}
\end{figure}

Finally, Fig. \ref{fig:temp-rv-uncertainty} shows that uncertainty for temperature extremes is very well-constrained, much more so than for precipitation and consistent with estimation of bounded distributions. The ensemble size of 10560 achieves relative uncertainty of less than 5\% in almost all cases.

\section{Discussion}

Our results suggest that statistical extreme value analysis can be used with huge climate emulator ensembles to estimate very low probability extreme precipitation and temperature, with well-constrained uncertainty and without the need for seasonal analysis or storm typing.

Our results for precipitation showing sensitivity to the choice of threshold, with declining shape parameter estimates with higher thresholds, are consistent with the results of \cite{ben2020evaluation}, who analyzed 1750 years of output from a regional climate model, finding decreasing shape parameter estimates with longer GEV block sizes and biased estimates when using annual maxima. 

A model- (i.e., simulation-) based approach carries numerous benefits. A central one is the large set of samples that characterize variability, with each sample capturing realistic spatio-temporal structure. One can calculate any function of interest (such as total precipitation over a particular watershed and time duration) for each sample. The resulting samples of the function characterize the variability of that quantity, with no need for complicated statistical techniques to account for joint uncertainty and the effects of nonlinearity. 

The major limitation is that the model/emulator must be fit for purpose, producing output similar enough to the real world for the estimates to be useful. This requires very careful assessment, assessment that is not part of this work and which for the most extreme events is by definition difficult or impossible. An additional limitation is the need for higher-resolution output in space and time, both for understanding impacts of interest and potentially for validating results from lower-resolution output, with sub-daily extremes being particularly challenging \citep{nasem2024pmp}. Finally, we note that the forcing conditions (from the years 2001-2022), particularly the SSTs, may not be sufficiently representative and may constrain the simulated weather produced.

One qualitative opportunity for assessment and improved scientific understanding  is to examine the meteorology associated with very extreme simulated events (e.g., the likely presence of extreme precipitation in southern Nevada indicated by the high 1-in-100000 year AEP depths in Fig.\ \ref{fig:prec-climate}) to assess plausibility and investigate potential mechanisms for events never seen before. Another opportunity is to use the estimates to quantify the likelihood of very extreme events in the historical record.

An additional limitation of using a reanalysis-based product and assuming stationarity regards climate change. Emulators trained on a specific time period may not generalize to conditions not present in the training sample. One can use emulators trained on climate models \citep[e.g.,][]{watt2025ace2}, but this introduces additional questions about fitness for purpose. To account for nonstationarity, EVA has a long history of using time or proxies for global warming as covariates \citep{coles2001introduction,westra2013global,risser2019probabilistic}. Alternatively, one could run an emulator under multiple climate scenarios and estimate AEP depths, either separately or jointly, to assess the effects of climate change. We see the main challenges as developing an appropriate emulator and deciding on the scenarios, not in the statistical analysis.

Much larger sample sizes than 10560 will likely be available in the near future, raising the possibility of simply using empirical quantiles rather than EVA. However, empirical quantiles are also statistical estimates with associated uncertainty, and improvements in emulator efficiency will likely be leveraged at least in part towards emulation at higher resolution or for multiple scenarios, suggesting a need for EVA for the foreseeable future.

\clearpage
\acknowledgments
We thank the Allen Institute/ACE2 project for providing the trained emulator, initial conditions and forcing data, and emulator code. We thank the Statistical Computing Facility at UC Berkeley, in particular GPUs provided by Jacob Steinhardt, for computing resources. We thank the  NASEM committee on modernizing PMP for inspiration. This research was supported by the Director, Office of Science, Office of Biological and Environmental Research of the U.S. Department of Energy under Contract DE-AC02-05CH11231 and by NSF grant DMS-2311164.

\datastatement
All code for creating the ensemble and analyzing the output is available on GitHub at \url{https://github.com/paciorek/huge-ensemble-eva}.
All input data for the emulator are also available, as summarized in Appendix A, with more details in the GitHub repository above.

\bibliographystyle{ametsocV6}
\bibliography{refs}

    
\appendix[A] 
\appendixtitle{Methods details}
\label{app-methods}

\subsection*{Ensemble design}
\label{app:design}


To generate the ensemble, we used the version of the model provided at \url{https://github.com/ai2cm/ace} (commit 2dceb9544c3092501f30ea16c69d8adb496d88c1) with the model ``checkpoint" (trained model parameters) and forcing and initial conditions files provided at \url{https://huggingface.co/allenai/ACE2-ERA5}. The checkpoint uses ERA5 training data from the years 1940–1995 and 2011–2019. For our ensemble, we used forcing data from 2001-2022 and initial conditions from 12 days in 2001 (the first day of each month). For each of the 12 sets of initial conditions, we ran the emulator forward over 40 repetitions of the forcing data plus one additional emulation year for 2001 to allow us to discard the data from the initial partial year 2001. In other words, when the emulator reached December 31, 2022, the next day of forcing data was January 1, 2001. While this recycling introduces a discontinuity in forcing variables, given the relatively short 22-year period, we expect this discontinuity to be moderate and not to have substantive impact on the simulated weather. A run of 40 repetitions of 22 years took approximately 36 hours on a single A100 GPU.  For annual maxima analysis, we removed leap years for data processing convenience.

\subsection*{Statistical extreme value analysis}
\label{app:eva}

As described in detail in \citet{coles2001introduction}, classic extreme value results characterize the limiting distribution of renormalized block maxima as sample size increases.
That is, given $d$ iid copies $X_{1}, \ldots, X_{d}$ of a random variable $X$ and letting $M_d = \max(X_1, \ldots, X_d)$, if there exist renormalizing sequences $a_d$ and $b_d$ such that $(M_d - a_d)/b_d$ converges as $d$ increases, it must converge to one of three types of extreme value distributions, which are encompassed by the generalized extreme value (GEV) distribution.
If $d$ is large, 
\begin{equation}
\label{eq:GEV}
P(M_d \leq x) \approx \exp
  \left\{ - \left[1 + \xi \left(\frac{x - \mu}{\sigma}\right) \right]^{-1/\xi} \right\},
\end{equation}
for $x$ such that $1 + \xi (x - \mu)/{\sigma} > 0$ and where the normalizing sequences have been absorbed into the location parameter, $\mu$, and scale parameter, $\sigma$. 
The shape parameter, $\xi$, is critical to understanding the nature of very low probability values.
If $\xi <0$, the upper tail is bounded. 
If $\xi = 0$, the upper tail is unbounded but `light' and all moments are finite.
If $\xi > 0$, the unbounded upper tail is heavy, with only moments less than $1/\xi$ being finite.
In practice, beginning with $n$ blocks of data each of size $d$,  one obtains the sample block maxima, $m_i = \max(x_{i,1}, \ldots, x_{i,d}), i = 1, \ldots, n$, and uses these data to estimate the three parameters of the GEV distribution.
With these parameter estimates, quantities of interest, such as the quantile corresponding to the 1-in-$T$ AEP depth, can be estimated with corresponding uncertainty estimates.

Similar results show that if a normalizing function $\sigma(u)$ can be found such that the distribution of $[(X - u)/\sigma(u) \mid X > u]$ (i.e., the conditional distribution conditioning on the observation exceeding $u$) converges as the threshold, $u$, increases, then this distribution converges to the generalized Pareto distribution (GPD).
For $u$ large, 
\begin{equation}
\label{eq:GPD}
P(X \leq x \mid X > u) \approx 1 - \left[1 + \xi \left( \frac{x - u}{\sigma'} \right) \right]^{-1/\xi},
\end{equation}
for $x$ such that $1 + \xi ( x - u)/\sigma > 0$, where the shape parameter, $\xi$, indexes the tail behavior as with the GEV, and the scale parameter, $\sigma'$, depends on the threshold $u$.
In practice, given a data set of $N$ iid observations, $x_i, i = 1, \ldots, N$, a threshold $u$ is chosen, and the threshold exceedances $x_{(1)}, \ldots, x_{(n)} > u, n < N$ are used to estimate the GPD parameters, $\xi$ and $\sigma'$. In this work, we use a point process representation of the threshold exceedance approach \citep[Section 7]{smith1989extreme,coles2001introduction}, with the parameters of the representation being equivalent to the three parameters of the GEV distribution.

\clearpage
\appendix[B]
\label{app-figs}
\appendixtitle{Additional figures}

\subsection*{Precipitation}

\begin{figure}[H]
    \centering
    \includegraphics[width=\linewidth]{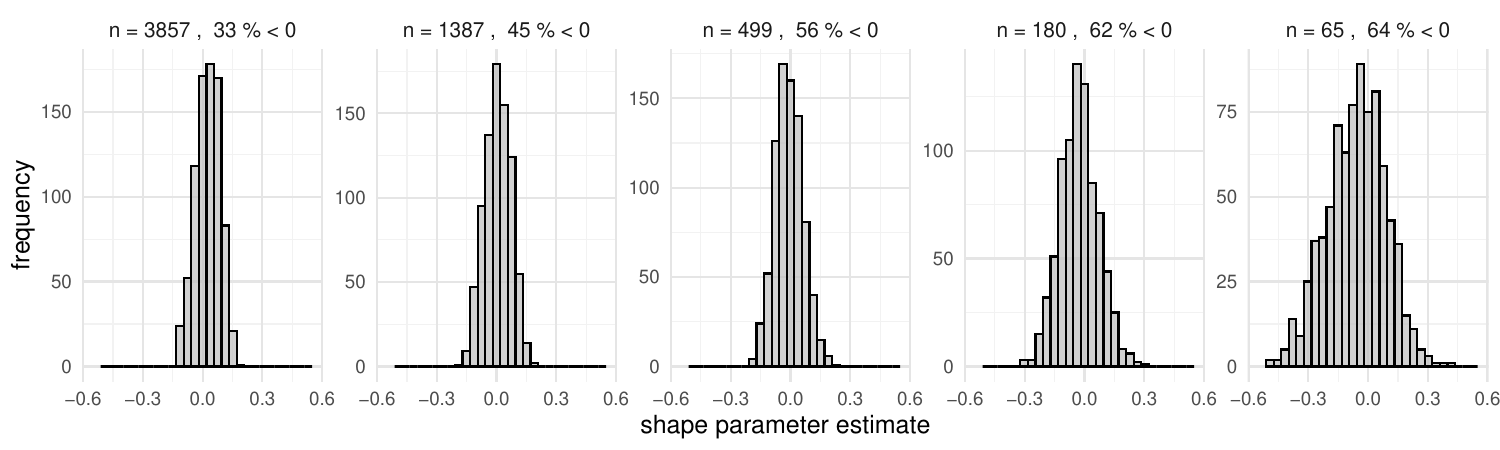}
    \caption{Change in distributions of precipitation shape parameter estimates with threshold for POT-based estimates.}
    \label{fig:prec-shape-hists}
\end{figure}

Fig. \ref{fig:prec-shape-hists} shows
that the number of positive shape parameter estimates decreases substantially as the POT threshold increases, with the distributions shifting to the left (as well as becoming more variable with smaller sample sizes)

\begin{figure}[H]
    \centering
    \includegraphics[width=\linewidth]{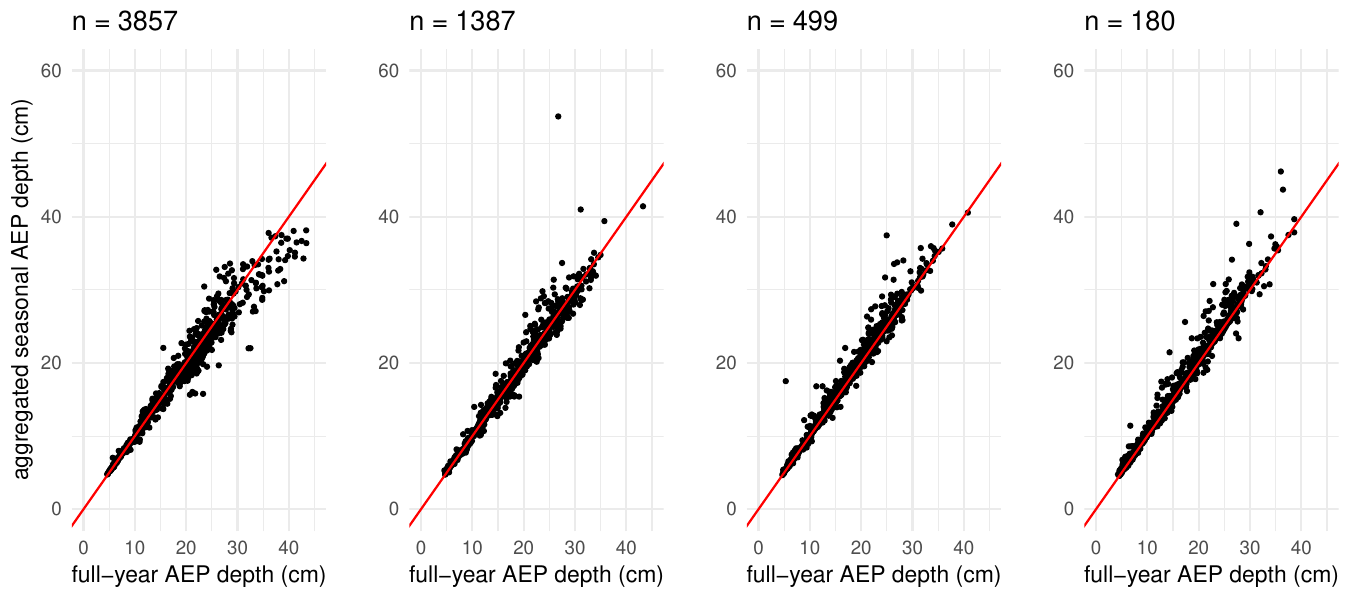}
    \caption{Comparison of 1-in-100000 year AEP precipitation depth estimates based on analysis of full-year data to aggregation over seasonally-stratified estimates. Results are shown for the first approach discussed in the methods. One-to-one lines are in red.}
    \label{fig:prec-rv-seasonal1}
\end{figure}

Fig. \ref{fig:prec-rv-seasonal1} shows that if one fits season-specific POT models and then determines the overall AEP depth as the largest amongst the four seasons using the first approach described in the Methods, the results are very similar to ignoring seasonality, provided the threshold is sufficiently large.

\subsection*{Temperature}

\begin{figure}[H]
    \centering
    \includegraphics[width=\linewidth]{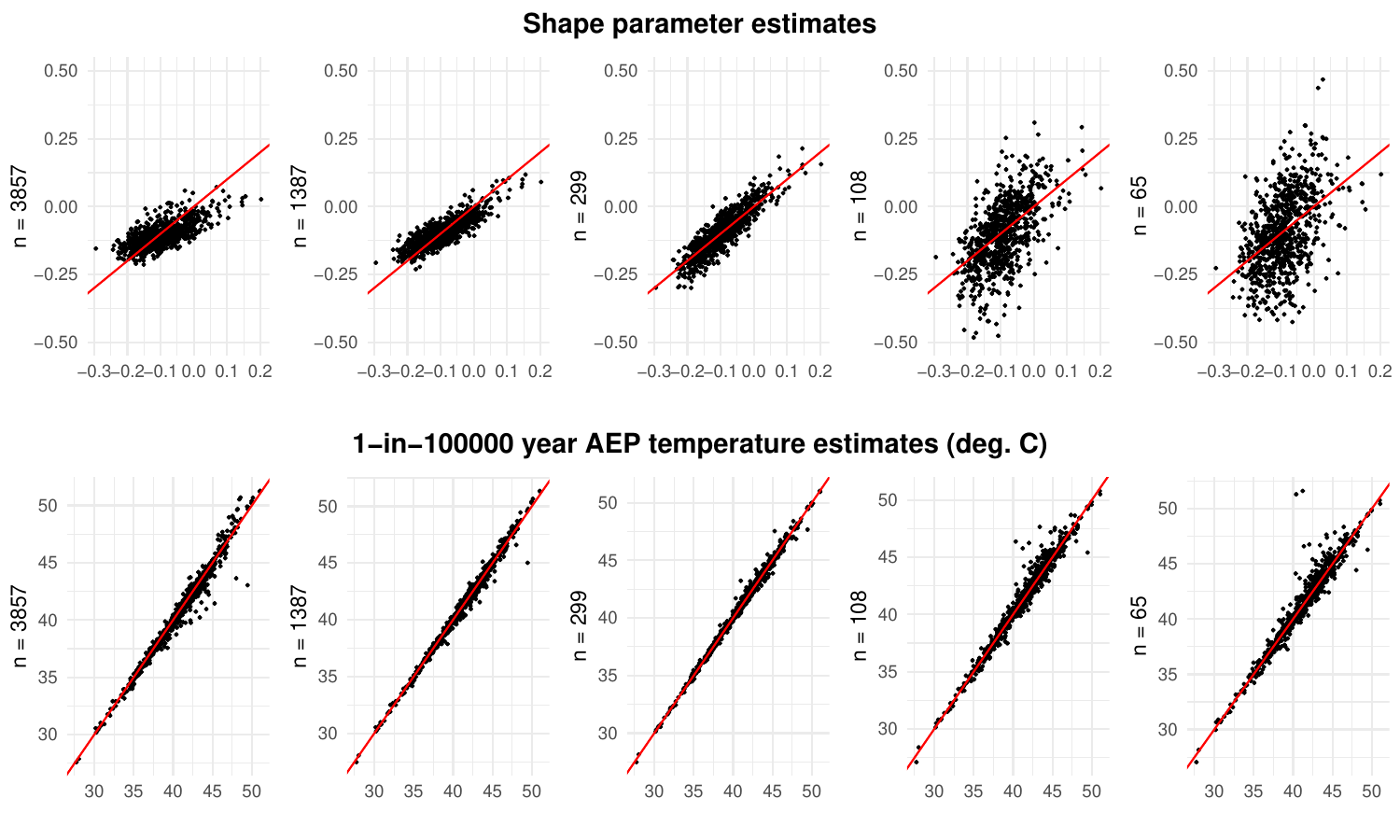}
    \caption{Change in temperature estimates with threshold for POT-based estimates. (Top row) Shape parameter estimates with increasing threshold on y-axis, compared to estimates with $n=499$ on x-axis. (Bottom row) 1-in-100000 year AEP temperature estimates with increasing threshold, compared to estimates with $n=499$. One-to-one lines are in red.}
    \label{fig:temp-stability}
\end{figure}

Fig. \ref{fig:temp-stability} shows that AEP temperature estimation is robust to the choice of threshold in the POT analysis, even for low thresholds (unlike for precipitation).

\begin{figure}[H]
    \centering
    \includegraphics[width=\linewidth]{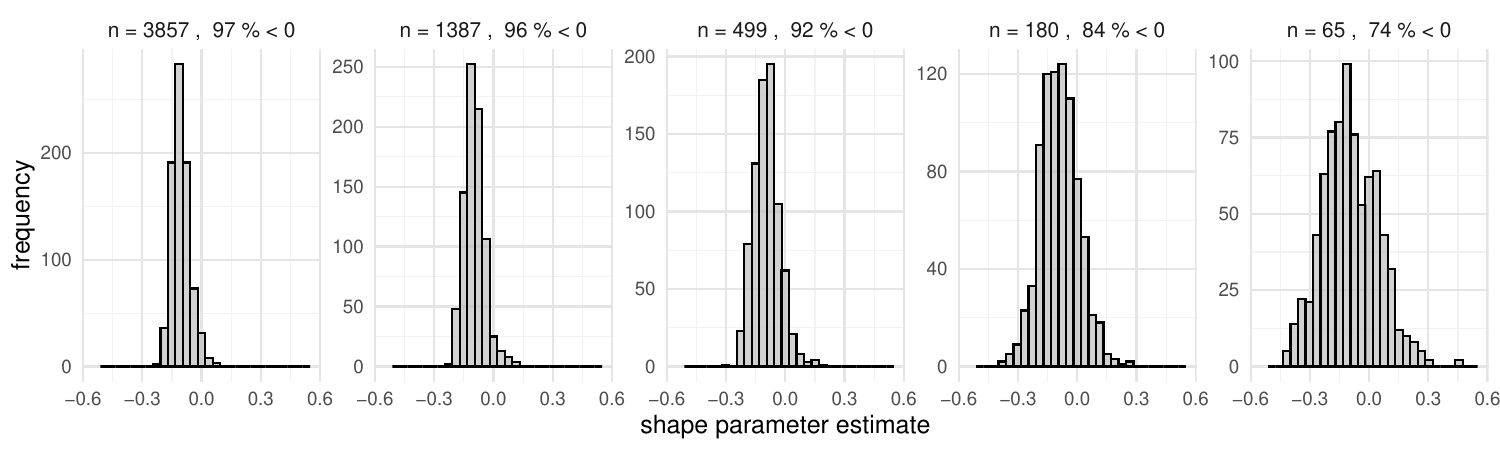}
    \caption{Change in distributions of temperature shape parameter estimates with threshold for POT-based estimates.}
    \label{fig:temp-shape-hists}
\end{figure}

Fig. \ref{fig:temp-shape-hists} shows that shape parameter estimates for temperature do not show a relationship with the threshold (apart from increasing variability with smaller sample sizes), unlike for precipitation.

\begin{figure}[H]
    \centering
    \includegraphics[width=\linewidth]{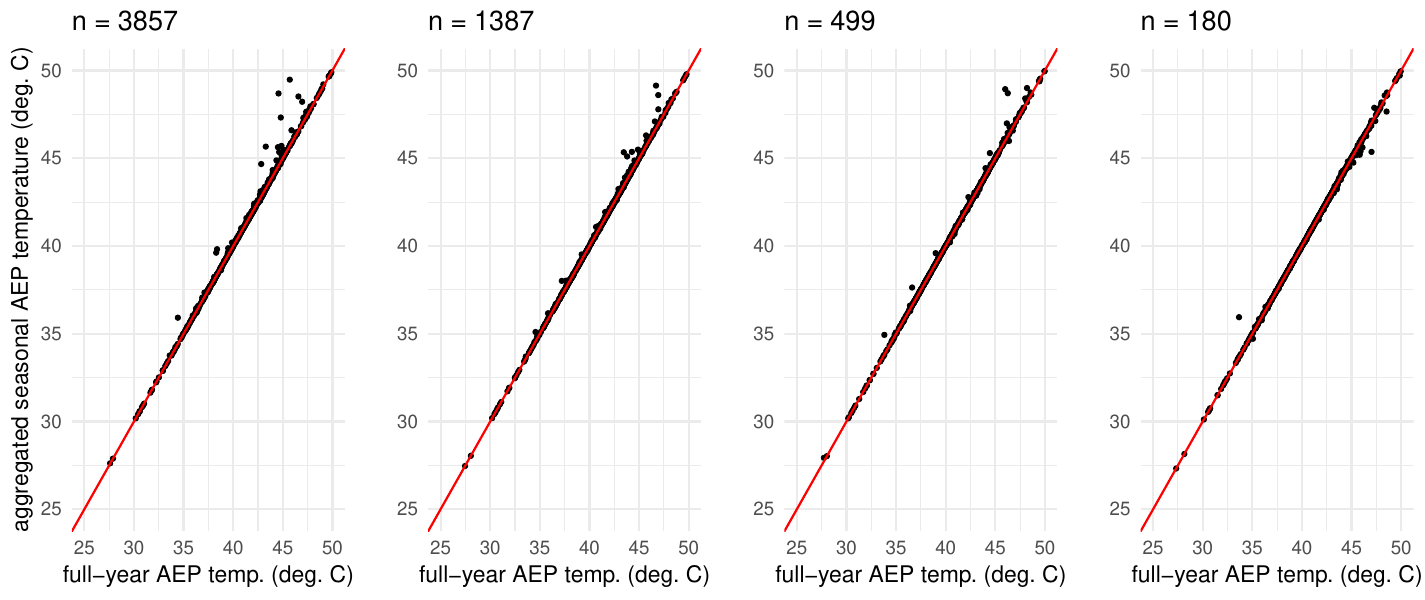}
    \caption{Comparison of 1-in-100000 year AEP temperature estimates based on analysis of full-year data to aggregation over seasonally-stratified estimates. Results are shown for the first approach discussed in the methods. One-to-one lines are in red.}
    \label{fig:temp-rv-seasonal1}
\end{figure}

Fig. \ref{fig:temp-rv-seasonal1} shows that if one fits season-specific POT models and then determines the AEP temperature as the largest amongst the four seasons using the first approach described in the Methods, the results are almost identical.

\end{document}